\documentclass[11pt,a4paper]{article} 
\pdfoutput=1
\usepackage{jcappub}
\usepackage{latexsym,amsmath,amssymb}
\usepackage{graphicx}
\usepackage{slashed}  
\usepackage{bm}
\usepackage{epsfig}
\usepackage{dcolumn}

\usepackage{bm}
\usepackage{latexsym}
\usepackage{amsmath}
\numberwithin{equation}{section}
\def\ee{\end{equation}}
\def\ba{\begin{eqnarray}}
\def\ea{\end{eqnarray}}

\def\bq{\begin{quote}}
\def\eq{\end{quote}}

 at 10truept

\newcommand{\beq}{\begin{equation}}
\newcommand{\eeq}{\end{equation}}
\newcommand{\beqa}{\begin{eqnarray}}
\newcommand{\eeqa}{\end{eqnarray}}
\newcommand{\bea}{\begin{eqnarray}}
\newcommand{\eea}{\end{eqnarray}}

 \newcommand{\ep}{\epsilon}

\newcommand{\Mpc}{\text{Mpc}}

\def\baq{\begin{eqnarray}}
\def\eaq{\end{eqnarray}}

\def\pa{\partial}

\def\lesssim{~\mbox{\raisebox{-.6ex}{$\stackrel{<}{\sim}$}}~}

\def\ltap{\ \raise.3ex\hbox{$<$\kern-.75em\lower1ex\hbox{$\sim$}}\ }
\def\gtap{\ \raise.3ex\hbox{$>$\kern-.75em\lower1ex\hbox{$\sim$}}\ }
\def\gl{\ \raise.5ex\hbox{$>$}\kern-.8em\lower.5ex\hbox{$<$}\ }
\def\roughly#1{\raise.3ex\hbox{$#1$\kern-.75em\lower1ex\hbox{$\sim$}}}

\title{\begin{center} Inflationary Magnetogenesis \\ without the Strong Coupling Problem \end{center} }
\subheader{DNRF90}

\author{Ricardo J. Z. Ferreira,}
\author{Rajeev Kumar Jain}
\author{and Martin S. Sloth}

\affiliation{CP$^3$-Origins, Centre for Cosmology and Particle Physics Phenomenology, 
University of Southern Denmark, Campusvej 55, 5230 Odense M, Denmark}

\emailAdd{ferreira@cp3.dias.sdu.dk}
\emailAdd{jain@cp3.dias.sdu.dk}
\emailAdd{sloth@cp3.dias.sdu.dk}

\date{\today}

\abstract{
The simplest gauge invariant models of inflationary magnetogenesis are known to suffer from the problems of either large backreaction or strong coupling, which make it difficult to self-consistently achieve cosmic magnetic fields from inflation with a field strength larger than $10^{-32} G$ today on the $\Mpc$ scale. Such a strength is insufficient to act as seed for the galactic dynamo effect, which requires a magnetic field larger than $10^{-20} G$. In this paper we analyze simple extensions of the minimal model, which avoid both the strong coupling and back reaction problems, in order to generate sufficiently large magnetic fields on the Mpc scale today. First we study the possibility that the coupling function which breaks the conformal invariance of electromagnetism is non-monotonic with sharp features. Subsequently, we consider the effect of lowering the energy scale of inflation jointly with a scenario of prolonged reheating where the universe is dominated by a stiff fluid for a short period after inflation. In the latter case, a systematic study shows upper bounds for the magnetic field strength today on the Mpc scale of $10^{-13} G$ for low scale inflation and $10^{-25} G$ for high scale inflation, thus improving on the previous result by 7-19 orders of magnitude. These results are consistent with the strong coupling and backreaction constraints.}
\keywords{Inflation, primordial magnetic fields}
\arxivnumber{}

\begin{document}
\maketitle

\section{Introduction}

Various astrophysical observations indicate that our universe is magnetized on different length scales~\cite{Grasso:2000wj,Kronberg:1993vk,Widrow:2002ud}. The coherent magnetic fields are not only present in bound cosmological structures e.g. stars, galaxies and cluster of galaxies but they also seem to be present in the intergalactic medium. While the typical field strength in galaxies and clusters is of the order of a few micro-Gauss \cite{Clarke:2000bz}, a careful study of some astrophysical processes seems to suggest a lower bound of a few femto-Gauss on the coherent magnetic fields in the intergalactic medium \cite{Neronov:1900zz}(for some recent reviews, see also \cite{Kandus:2010nw,Widrow:2011hs,Ryu:2011hu,Durrer:2013pga}).

The origin of these magnetic fields has not yet been completely understood. Various explanations that have been put forward can be broadly classified into two categories: Primordial and Astrophysical. According to the primordial hypothesis, large scale magnetic fields are either created during an inflationary phase or during the primeval phase transitions (electroweak or QCD) in the early universe. These initial seed fields are then amplified by the dynamo action at later epochs to lead to the observed strength \cite{Brandenburg:2004jv}. On the other hand, the astrophysical process presumes that the seed fields are indeed generated by the plasma effects and then boosted up by the dynamo mechanism which can possibly explain the field strength in galaxies and also in clusters having typical coherence length of $0.01-1$ \Mpc. However, the recently claimed (indirect) detection of large scale coherent magnetic fields in cosmic voids seems very difficult to explain by astrophysical processes, which suggests that such fields could have a primordial origin coming from the early universe.

Among the primordial mechanisms, magnetic fields generated during phase transitions can have relevant strength but the correlation lengths are typically too small to explain cosmic magnetic fields \cite{Durrer:2003ja}. On the contrary, inflation produced magnetic fields are very interesting due to large correlation scale as well as pertinent field strengths. Due to the fact that the Electromagnetic (EM) Lagrangian is conformally invariant in the Friedmann-Lema\^{\i}tre-Robertson-Walker (FLRW) background, the EM field is not amplified during the expansion of the universe. Therefore, in order to generate magnetic fields during inflation, a necessary condition is to break the conformal invariance either by coupling the EM field to a time dependent background field (inflaton, curvaton, etc.) or by introducing another coupling which breaks conformal invariance or even gauge invariance. Such possibilities have been investigated in \cite{Turner:1987bw, Ratra:1991bn}  and later have been revisited in \cite{Martin:2007ue,Giovannini:2007rh,Subramanian:2009fu,Kanno:2009ei}. Here, we will consider the simplest gauge invariant possibility of a coupling between the EM field and the inflaton $\phi$ parametrized as $f^2(\phi)F_{\mu\nu}F^{\mu\nu}$ where $F_{\mu\nu}$ is the EM field tensor and defined as $F_{\mu\nu} \equiv \pa_\mu A_\nu-\pa_\nu A_\mu$. The time dependence of such a coupling through the inflaton leads to the excitations of the EM field fluctuations and generate magnetic fields\footnote{Apart from the normal magnetic fields, generation of helical magnetic fields have also been studied during inflation. In such models, the final strength of magnetic fields typically turns out to be very small and in some cases, strong backreaction is unavoidable \cite{Anber:2006xt,Campanelli:2008kh,Durrer:2010mq,Byrnes:2011aa,Jain:2012jy}.}. At the end of inflation, it is assumed that classical electromagnetism is restored and therefore, $f\left(\phi\right)\rightarrow1$. Although this class of models can generate relevant magnetic fields at the present epoch with an appropriate choice of the coupling function, they indeed suffer from some severe problems. The first is the so-called backreaction problem wherein the energy density of the EM field spoils the inflationary background dynamics which can be avoided by a suitable choice of parameters of the model. The second is the strong coupling problem in which the time evolution of the coupling during inflation leads to the EM field being in the strong coupling regime at the beginning of inflation so the perturbative calculation of the EM field fluctuations can not be trusted, as pointed out by Demozzi, Mukhanov and Rubinstein \cite{Demozzi:2009fu}. Interestingly, this class of models have existed in the literature for a long time, the strong coupling problem was only noticed relatively recently. Due to these problems, it has been realized that constructing a realistic self-consistent model of magnetic field generation during inflation is quite difficult and even upper limits on the present magnetic fields have been derived \cite{Fujita:2012rb}.

Recently, such couplings have also been considered beyond the context of inflationary magnetogenesis. For instance, a time-dependent interaction between the inflaton and the vector fields can induce non-Gaussian cross-correlations between the metric/curvature perturbations and magnetic fields which turn out to be large for a particular shape and could have interesting cosmological consequences \cite{Caldwell:2011ra,Motta:2012rn,Barnaby:2012tk,Jain:2012ga,Jain:2012vm,Shiraishi:2012xt}.  Furthermore, statistically anisotropic contribution to the primordial curvature perturbation during inflation as well as anisotropic power spectrum and bispectrum due to the presence of this coupling have also been explored \cite{Watanabe:2010fh,Bartolo:2012sd,Lyth:2013sha,Urban:2013aka}.

\subsection{Summary of the proposed models}

In this subsection we will briefly summarize the two different studies which have been done in this paper.

In both approaches we start from the assumption that the conformal invariance of the EM action has to be broken. Let us consider the coupling of the inflaton, $\phi$, to the EM field given as above by $f^2(\phi)F_{\mu\nu}F^{\mu\nu}$. The inflaton field will be a slowly varying function of the scale factor, $a(t)$, and therefore, $f(\phi)$ will also depend on the scale factor which in turn breaks the conformal invariance of the EM action. 
For simplicity, if we assume that the coupling function is a simple monotonic monomial function of the scale factor as $f \propto a^{\alpha}$, it is well known that in the two cases of either a growing ($\alpha = 2$) or a decreasing ($\alpha =-3$) coupling function, one can obtain a scale invariant spectrum of cosmic magnetic fields from inflation with $d\rho_B/d\log k \approx H^4$, where $H$ is the Hubble scale during inflation. With $H = 10^{-6} M_{p}$, this corresponds to a magnetic field strength of order $10^{-12} G$ on large scales today. Here $\rho_B$ denotes the energy density of the magnetic field  and $M_{p}$ is the reduced Planck mass. However, the mechanism can not work in this simple form. Given the assumption that the coupling function is a simple monotonic monomial function of the scale factor, it was shown by Demozzi, Mukhanov and Rubinstein that either of the two problems will occur \cite{Demozzi:2009fu}.

If the coupling function is decreasing with $\alpha = -3$, the energy density in the electric field grows rapidly, and after a little more than $10$ e-folds of inflation, it becomes larger than the background energy density $\rho_c = 3H^2 M_{p}^2$ leading to a backreaction problem. 
On the other hand, for the case of an increasing coupling function with $\alpha =2$, there is no backreaction problem but instead one encounters the strong coupling problem. If the coupling function is increasing, it means that it must have been very small in the past. If the EM field is coupled to charged matter in the usual gauge invariant manner then rewriting the Lagrangian in terms of the canonically normalized EM field, the physical coupling will scale as $e_{\rm phys} = e/f$. Assuming that $f \rightarrow1$ at the end of inflation, $e_{\rm phys}$ will have to be very large shortly into the inflationary regime, and our ability to make trustable predictions during inflation will break down.

In order to balance between the backreaction and the strong coupling problems, we are therefore lead in our first approach to dispose of the assumption that $f$ is a simple monotonic monomial function of the scale factor. The idea is to construct a coupling function consisting of piecewise sections with different slopes, as the figurative example in Figure \ref{fig:Coupling function}. 
We considered two cases, one with just one transition and other with two. In both cases our coupling function was constructed in such a way that both back reaction and strong coupling problems are avoided. However, as we shall explain in Section \ref{sec:ourmodel}, by using the appropriate matching conditions, the dominant solution before the transition matches to the decaying solution after the transition. This leads to a very large $k$-dependent loss in the magnetic field spectrum in all the transitions from $\alpha>-1/2$ to $\alpha<-1/2$ and in the electric field spectrum for the opposite cases. We searched for regions where the magnetic fields were more enhanced than the standard situation without transitions. However, only in the case of one transition we get an improvement and of no more than 1 order of magnitude.

Instead, we will then subsequently study another possibility for obtaining large magnetic fields from inflation. While it is possible to have large magnetic fields at the end of inflation, the problem is that the magnetic field decays as $1/a^2(t)$ during all of the subsequent history of the universe due to magnetic flux conservation. This implies that the large magnetic fields generated during inflation are completely diluted. This also implies that the strength of magnetic fields today is very sensitive to the amount of redshift, which takes place after inflation. In order to get the maximum out of the magnetic field created from inflation, and derive an upper bound on the field strength today of magnetic fields from inflation, we should consider the case with a minimum of redshift after inflation.

The amount of redshift taking place after inflation is determined by the energy scale of inflation and the equation of state after inflation. The faster the energy density redshifts with the scale factor, the less the scale factor needs to increase after inflation for the energy density to reach the value it has today. Thus, we find that the optimal case is when the energy scale of inflation is as low as possible and the universe is dominated by a stiff fluid after inflation (called deflation \cite{Spokoiny:1993kt}, like, for example, in disformal \cite{Kaloper:2003yf} or quintessence inflation \cite{Peebles:1998qn,Giovannini:2003jw}) where the energy density redshifts rapidly as $1/a^6(t)$ until shortly before Big-Bang nucleosynthesis, when reheating is assumed to take place. In this scenario, we derive an equation which gives the maximal magnetic field today as a function of the Hubble parameter during inflation\footnote{Some related works considering the effects of the effective reheating parameter are \cite{Martin:2007ue,Demozzi:2012wh}.}. From it one can show that the standard mechanism of breaking conformal invariance during inflation can generate large scales magnetic fields as large as $10^{-13} G$. This is the main result of this paper.

\subsection{Outline of the paper}

This article has the following structure. In the next section, we briefly present the essentials of inflationary magnetogenesis by introducing a time dependent coupling function for the EM action which breaks conformal invariance and derive the electric and magnetic power spectrum for a given coupling function. In Section \ref{sec:scp}, we explain the strong coupling problem by computing explicitly the EM power spectrum and backreaction for a coupling function having a power law behavior in the scale factor/conformal time. Section \ref{sec:ourmodel} shows the effect of having transitions in the coupling function on the magnetic fields and its spectrum. In Section \ref{sec:reh}, we perform an analysis of the effect on the present magnetic fields of lowering the scale of inflation and of prolonged reheating.  Finally, in Section \ref{sec:conclusion}, we discuss our results and conclude with a few comments.

Throughout this paper, we work in natural units with $\hbar=c=1$, and the reduced Planck mass $M_{p}^2\equiv 1/8 \pi G$ is set to unity except at a few places. Our metric signature is $(-,+,+,+)$. We use Greek indices $\mu, \nu, ...$ etc. for space time coordinates and Latin indices $i, j, ...$ for spatial coordinates.

\section{Essentials of inflationary magnetogenesis} \label{sec:infmf}

In this section, we quickly review the key details of the mechanism of inflationary magnetogenesis. Let us start with the standard EM action, given by
\begin{equation}
S_{EM}=-\frac{1}{4}\int d^{4}x\, \sqrt{-g}\, F_{\mu\nu}F^{\mu\nu}.\label{eq:Electromagnetic Lagrangian}
\end{equation}
It is well known that this action is conformally invariant in a FLRW space time and therefore, one can not amplify the EM field fluctuations which leads to an adiabatic decay of EM field as $1/a^2$ with the expansion of the universe.
Inflationary mechanisms of magnetic field generation therefore require the breaking of conformal invariance of EM action. A large number of possibilities have been considered for this purpose. The simplest (gauge invariant) of these possibilities is to introduce a time dependent coupling as
\begin{equation}
S=-\frac{1}{4}\int d^{4}x\, \sqrt{-g}\, f^{2}\left(\phi\right)F_{\mu\nu}F^{\mu\nu},\label{eq:Electromagnetic Action}
\end{equation}
where $\phi$ is a homogeneous scalar field which is assumed to be the inflaton.  For a flat FLRW space time given by
\begin{equation}
ds^{2}=a^2(\eta)\left(-d\eta^{2}+dx^{2}+dy^{2}+dz^{2}\right),\label{eq:FRW space time}
\end{equation}
where $\eta$ is the conformal time, and working in the Coulomb gauge
where $A_{0}=0$ and $\partial_{i}A^{i}=0$ Eq. (\ref{eq:Electromagnetic Action}) yields
\begin{equation}
S=\frac{1}{2}\int d^{3}x\, d\eta\, f^{2}\left(\phi \right)\left(A_{i}'^{2}-\frac{1}{2}\left(\partial_{i}A_{j}-\partial_{j}A_{i}\right)^{2}\right),\label{eq:Electromagnetic Action in FRW and Coulomb Gauge}
\end{equation}
where $'\equiv\frac{d}{d\eta}$. From Eq. (\ref{eq:Electromagnetic Action in FRW and Coulomb Gauge})
we obtain the following equation of motion
\begin{equation}
A_{i}''+2\frac{f'}{f}A_{i}'-a^{2}\partial_{j}\partial^{j}A_{i}=0.\label{eq:Equation of motion in position space}
\end{equation}
In order to quantize the EM field, we promote the vector field to an operator and perform a Fourier expansion as follows

\begin{equation}
A_{i}\left(\eta,\mathbf{x}\right)=\sum_{\sigma=1,2}\int\frac{d^{3}k}{\left(2\pi\right)^{3/2}}\left(\epsilon_{i,\sigma}\left(\mathbf{k}\right)A_{\mathbf{k}}\left(\eta\right)\hat{b}_{\mathbf{k,\sigma}}e^{i\mathbf{k}\mathbf{x}}+h.c.\right),\label{eq:Fourier expansion}
\end{equation}

where by definition $\delta_{ij}{\epsilon}^{i}{\epsilon}^{j}=1$
and the following identities are verified: 
\[{\epsilon}_{\sigma}^{i}k_{i}=0, \quad \sum_{\sigma=1,2}{\epsilon}_{\sigma}^{i}\left(\mathbf{k}\right){\epsilon}_{j,\sigma}\left(\mathbf{k}\right)=\delta_{j}^{i}- \delta_{jl}k^{i}k^{l}/{k^{2}}.\]
This allows us to impose the usual commutation relations
\begin{equation}
\left[\hat{b}_{\mathbf{k_{1}}}^{\sigma},\hat{b}_{\mathbf{k_{2}}}^{\sigma'\dagger}\right]=\left(2\pi\right)^{3}\delta^{\left(3\right)}\left(\mathbf{k_{1}}-\mathbf{k_{2}}\right) \delta_{\sigma \sigma'}.\label{eq:Commutation Relation}
\end{equation}
It is easy to see that the canonically normalized vector field associated with $A_{k}$ is ${\cal A}_{k}=f(\eta)A_{k}(\eta)$.
In terms of ${\cal A}_{k}$, Eq. (\ref{eq:Equation of motion in position space}) simplifies to
\begin{equation}
{\cal A}_k''+\left(k^{2}-\frac{f''}{f}\right){\cal A}_{k}=0.\label{eq:Simplified Equation of motion for generic coupling function}
\end{equation}
This mode equation for $\cal A$ resembles the equation of a harmonic oscillator with a time dependent mass term. In the limit of large wavelengths, i.e. $k^{2}\ll f''/f$, and if there are no sudden changes in the behavior of $f$, this equation can be directly integrated for a generic coupling function and the solution reads
\begin{equation}
{\cal A}_{k}(\eta) \sim C_{1}f+C_{2}f\int\frac{d\eta}{f^{2}},\label{eq:Generic Solution of the eq. of motion in the limit of large wavelengths}
\end{equation}
where $C_{1}$ and $C_{2}$ are integration constants which are fixed by imposing vacuum initial conditions at early times.


In order to compute the EM spectrum, we first note that its energy-momentum tensor is given by \cite{Martin:2007ue}
\begin{equation}
T_{\mu\nu}=f^{2}\left(F_{\mu}^{\ \beta}F_{\nu\beta}-\frac{1}{4}g_{\mu\nu}F_{\alpha\beta}F^{\alpha\beta}\right).\label{eq:Energy-Momentum Tensor}
\end{equation}
The total energy density in the EM field can then be written as a sum of the electric and magnetic contributions \cite{Subramanian:2009fu}
\begin{equation}
\rho_{EM}=\rho_{E}+\rho_{B},\label{eq:Total Energy Density}
\end{equation}
where
\beqa
\rho_{E}=\left\langle 0\left|T_{\mu\nu}^{E}u^{\mu}u^{\nu}\right|0\right\rangle =\frac{1}{2}f^{2}E_{i}E^{i}\label{eq:ElectricEnergyDensity},\\
\rho_{B}=\left\langle 0\left|T_{\mu\nu}^{B}u^{\mu}u^{\nu}\right|0\right\rangle =\frac{1}{2}f^{2}B_{i}B^{i}\label{eq:MagneticEnergyDensity}.
\eeqa
Using the four velocity of the fundamental observer $u^{\mu}=(1/a,0,0,0)$, the spatial components of the four vectors $E_{\mu}$ and $B_{\mu}$ are given by
\beq
E_{i}=-\frac{1}{a}A_{i}'\, , \qquad B_{i}=\frac{1}{a}\epsilon_{ijk}\partial_{j}A_{k}
\eeq
while the time components vanish. Using Eqs. (\ref{eq:ElectricEnergyDensity}) and (\ref{eq:MagneticEnergyDensity}), one can compute the magnetic and electric field spectrum as
\beqa
\frac{d\rho_{B}}{d\log k}&=&\frac{1}{2\pi^{2}}\frac{k^{5}}{a^{4}}\left|{\cal A}\left(k,\eta\right)\right|^{2}\label{eq:Magnetic Spectrum},\\
\frac{d\rho_{E}}{d\log k}&=&\frac{f^{2}}{2\pi^{2}}\frac{k^{3}}{a^{4}}\left|\left(\frac{{\cal A}\left(k,\eta\right)}{f}\right)'\right|^{2}.\label{eq:Electric Spectrum}
\eeqa
One can go even further without specifying the coupling function. In the limit of large wavelengths, Eq. (\ref{eq:Generic Solution of the eq. of motion in the limit of large wavelengths}) gives the leading order solution to the mode equation (\ref{eq:Simplified Equation of motion for generic coupling function}). Therefore, in this limit, one can write the magnetic and electric field spectrum explicitly as a function of the coupling. Note that, if the leading term is proportional to $f$, we get $d\rho_{E}/d\log k \propto \left| \left({\cal A}/f\right)'\right|^2 = 0$, and in this case, one needs to go to the second order to obtain the non-vanishing electric spectrum which requires solving the full mode equation.

\section{Review of the strong coupling problem \label{sec:scp}}

In order to review the strong coupling problem, we shall consider a specific class of coupling function which has received great attention. In this case, the coupling function has a power law dependence on the scale factor as
\beq
f\left(\eta\right)\propto a^{\alpha},
\eeq
where $\alpha$ is a free parameter. As we shall discuss later, such a parametrization leads to interesting results for certain values of $\alpha$. In a de Sitter space time, the scale factor evolves as $a\left(\eta\right)=a_{0}\left|{\eta_0}/{\eta}\right|$, where the conformal time $\eta$ goes from $-\infty$ to $0_{-}$. With these assumptions, the Fourier mode equation becomes
\begin{equation}
{\cal A}_{k}''+\left(k^{2}-\frac{\alpha\left(\alpha+1\right)}{\eta^{2}}\right){\cal A}_{k}=0,\label{eq:Simplified equation of motion for a power law coupling}
\end{equation}
whose solution can be written in terms of Bessel functions as
\begin{equation}
{\cal A}_{k}=\left(-k\eta\right)^{1/2}\Big[C_{1}\left(k,\alpha\right)J_{-\alpha-1/2}\left(-k\eta\right)+C_{2}\left(k,\alpha\right)J_{1/2+\alpha}\left(-k\eta\right)\Big],\label{eq:Solution of the eq. of motion for a power law coupling}
\end{equation}
where by imposing the initial conditions ${\cal A}_{k}\rightarrow e^{-ik\eta}/\sqrt{2k}$ as $(-k \eta) \rightarrow \infty$, the two integration constants $C_1$ and $C_2$ are fixed to be
\beq
C_{1}\left(k,\alpha\right)=\sqrt{\frac{\pi}{4k}}\frac{\exp\left(i\pi\alpha/2\right)}{\cos\left(\pi\alpha\right)}, \qquad
C_{2}\left(k,\alpha\right)=\sqrt{\frac{\pi}{4k}}\frac{\exp\left(i\pi\left(1-\alpha\right)/2\right)}{\cos\left(\pi(1-\alpha)\right)}.
\eeq
In the late time limit or, equivalently, for modes well outside the Hubble radius $\left(-k\eta\rightarrow0\right)$, one can use the Taylor expansion of the Bessel functions around zero
\begin{equation}
J_{\nu}\left(x\right)=\sum_{m=0}^{\infty}\frac{\left(-1\right)^{m}}{m!\,\Gamma\left(m+\nu+1\right)}\left(\frac{x}{2}\right)^{2m+\nu},\label{eq:Bessel Function Taylor series}
\end{equation}
to obtain the leading terms of Eq. $\left(\ref{eq:Solution of the eq. of motion for a power law coupling}\right)$ as
\begin{equation}
{\cal A}_{k}\rightarrow\frac{2^{\alpha+1/2}\,C_{1}\left(k,\alpha\right)}{\Gamma\left(\frac{1}{2}-\alpha\right)}\left(-k\eta\right)^{-\alpha}+\frac{C_{2}\left(k,\alpha\right)}{2^{3/2+\alpha}\,\Gamma\left(\frac{3}{2}+\alpha\right)}\left(-k\eta\right)^{1+\alpha},\label{eq:Solution for the equation of motion at late times}
\end{equation}
at late times, in agreement with Eq. $\left(\ref{eq:Generic Solution of the eq. of motion in the limit of large wavelengths}\right)$. From Eq. (\ref{eq:Solution for the equation of motion at late times}), one can see that the first term dominates for $\alpha>-1/2$ while the second term dominates for $\alpha<-1/2$.

\subsection{The spectrum} \label{sec: The Spectrum}

Given the late time solution, we can now calculate the spectrum of magnetic and electric fields. Substituting the leading term of Eq. (\ref{eq:Solution for the equation of motion at late times})
into Eq. (\ref{eq:Magnetic Spectrum}) allow us to compute the magnetic spectrum
\begin{equation}
\frac{d\rho_{B}}{d\log k}\approx\frac{{\cal F}\left(n\right)}{2\pi^{2}}H^{4}\left(-k\eta\right)^{4+2n},\label{eq:Magnetic Spectrum Explicit}
\end{equation}
where 
\begin{equation}
n=\begin{cases}
\begin{array}{c}
-\alpha,\ \ \; \; \alpha\geq-1/2\\
1+\alpha,\ \alpha\leq-1/2
\end{array} & \mbox{\mbox{and}}\end{cases}{\cal \ F}\left(n\right)=\frac{\pi}{2^{2n+1}\Gamma^{2}\left(n+1/2\right)\cos^{2}\left(\pi n\right)},\label{eq:Constant definition for the magnetic spectrum}
\end{equation}
and we have used $aH=-1/\eta$ in a de Sitter background. From Eq. (\ref{eq:Magnetic Spectrum Explicit}), one can see that a scale invariant magnetic spectrum occurs for $\alpha=2$ or $\alpha=-3$. Notice that these two cases correspond to $\alpha\left(\alpha+1\right)=6$ which means that they satisfy exactly the same equation of motion and therefore, one cannot distinguish between them using the magnetic spectrum. One can also see that the magnetic spectrum at horizon exit for each mode is $\delta_{B}^{2}\equiv d\rho_{B}/d\log k\approx H^{4}$.

The same analysis can be done for the electric field, although there are a few differences. As mentioned at the end of Sec. \ref{sec:infmf}, when the first term of Eq. (\ref{eq:Solution for the equation of motion at late times}) dominates, $\left({\cal A}/f\right)'=0$ and hence, we have to go up to second order term in the Taylor expansion of Eq. (\ref{eq:Bessel Function Taylor series}) to write
\begin{equation}
{\cal A}\rightarrow C_{1}\left(k,\alpha\right)\left(-k\eta\right)^{-\alpha}\left(\frac{2^{\alpha+1/2}}{\Gamma\left(\frac{1}{2}-\alpha\right)}-\frac{\left(-k\eta\right)^{2}}{2^{3/2-\alpha}\Gamma\left(\frac{3}{2}-\alpha\right)}\right)+\frac{C_{2}\left(k,\alpha\right)}{2^{3/2+\alpha}\Gamma\left(\frac{3}{2}+\alpha\right)}\left(-k\eta\right)^{1+\alpha}.\label{eq:Solution for the eq. of motion up to 2nd order}
\end{equation}
When the second term of Eq. (\ref{eq:Solution for the equation of motion at late times}) is dominant, we do not need to consider the second order term. For all values of $\alpha$, the electric
spectrum can be written as
\begin{equation}
\frac{d\rho_{E}}{d\log k}\approx\frac{{\cal G}\left(m\right)}{2\pi^{2}}H^{4}\left(-k\eta\right)^{4+2m}\label{eq:Electric Spectrum Explicitly}
\end{equation}
where 
\begin{equation}
m=\begin{cases}
\begin{array}{c}
1-\alpha,\ \alpha \geq1/2\\
\alpha,\  \alpha \leq1/2
\end{array} & \mbox{\mbox{and}}\end{cases}{\cal \ G}\left(m\right)=\frac{\pi}{2^{2m+3}\Gamma^{2}\left(m+\frac{3}{2}\right)\cos^{2}\left(\pi m\right)}.\label{eq:Constants definition for the electric spectrum}
\end{equation}
From Eq. (\ref{eq:Electric Spectrum Explicitly}), one can see that the electric spectrum is scale invariant for $\alpha=3$ or $\alpha=-2$. In particular, let us analyse the case where the magnetic spectrum is scale invariant. For $\alpha=2$ the spectrum goes as $d\rho_{E}/d\log k\propto\left(-k\eta\right)^{2}$ which vanishes quickly as $\left(-k\eta\right)\rightarrow0$. On the other hand, when $\alpha=-3$ the spectrum goes as $d\rho_{E}/d\log k\propto\left(-k\eta\right)^{-2}$, which grows rapidly in the limit of large wavelengths.

One important point, which can be used as a consistency check of various results, is the duality between the electric and magnetic field. Namely, the electric field spectrum obtained from a given coupling function $f=a^{\alpha}$ is equivalent to the magnetic field spectrum obtained from $f=a^{-\alpha}$ (for a discussion of this duality, see Ref. \cite{Giovannini:2009xa}).

\subsection{The backreaction problem} \label{sec:backreaction}

In order to ensure that the the backreaction of the EM fields stays small during inflation, we shall first consider the energy stored in the electric field at a given
time $\eta_{f}=\left(a_{f}H\right)^{-1}$, which is given by
\[
\rho_{E}=
\int_{Ha_{i}}^{Ha_{f}}\frac{d\rho_{E}}{d\log k}d\log k=\frac{H^{4}}{2\pi^{2}}\int_{Ha_{i}}^{Ha_{f}}dk
\begin{cases}
\begin{array}{c}
4{\cal G}\left(1-\alpha\right)\left(-k\eta\right)^{6-2\alpha}k^{-1},\ \alpha\geq1/2\\
\left(1+2\alpha\right)^{2}{\cal G}\left(\alpha\right)\left(-k\eta\right)^{4+2\alpha}k^{-1},\ \alpha\leq1/2
\end{array}\end{cases}
\]

\begin{equation}
\approx\frac{H^{4}}{2\pi^{2}}\begin{cases}
\begin{array}{c}
4{\cal G}\left(1-\alpha\right)\begin{cases}
\begin{array}{c}
\frac{1}{6-2\alpha}\begin{cases}
\begin{array}{c}
1,\ \alpha<3\\
-\left(\frac{a_{i}}{a_{f}}\right)^{6-2\alpha},\ \alpha>3
\end{array}\end{cases}\\
\log\left(\frac{a_{f}}{a_{i}}\right),\ \alpha=3
\end{array} & ,\alpha\geq1/2\end{cases}\\
\left(1+2\alpha\right)^{2}{\cal G}\left(\alpha\right)\begin{cases}
\begin{array}{c}
\frac{1}{4+2\alpha}\begin{cases}
\begin{array}{c}
1,\ \alpha>-2\\
-\left(\frac{a_{i}}{a_{f}}\right)^{4+2\alpha},\ \alpha<-2
\end{array}\end{cases}\\
\log\left(\frac{a_{f}}{a_{i}}\right),\ \alpha=-2
\end{array} & ,\alpha\leq1/2\end{cases}
\end{array}\end{cases}\label{eq:electric energy density}
\end{equation}
where we have used Eqs. (\ref{eq:Electric Spectrum Explicitly}) and (\ref{eq:Constants definition for the electric spectrum}).

One can see that the case $\alpha=2$, where we have a magnetic scale invariant spectrum, do not lead to any backreaction. On the other hand, there are divergent behaviors, both for $\alpha>3$ and for $\alpha<-2$, where the energy rapidly grows. In order to avoid backreaction we have to ensure that $\rho_{EM}\lesssim H^{2}$. Considering $H\approx10^{-6}M_p$, similar to the example in Ref. \cite{Demozzi:2009fu}, one finds that the number of e-folds $N=\log\left(a_{f}/a_{i}\right)$
of inflation allowed for each of these two divergent regimes is 
\begin{equation}
N\lesssim\begin{cases}
\begin{array}{c}
\frac{-6}{3-\alpha}\log10,\ \alpha>3\\
\frac{-6}{\alpha+2}\log10,\ \alpha<-2
\end{array} & \end{cases}\label{eq:Allowed number of e-folds}
\end{equation}
For the other magnetic scale invariant
case with $\alpha=-3$, this implies that we cannot have more than approximately $13.8$ e-folds with this behavior. Therefore, inflation is not compatible with this specific coupling function. 

We now proceed to the energy density coming from the magnetic fields. Similarly, it yields
\[
\rho_{B}=\int_{Ha_{i}}^{Ha_{f}}\frac{d\rho_{B}}{d\log k}d\log k=\frac{H^{4}}{2\pi^{2}}\int_{Ha_{i}}^{Ha_{f}}dk\begin{cases}
\begin{array}{c}
{\cal F}\left(\alpha\right)\left(-k\eta\right)^{4-2\alpha}k^{-1},\ \alpha\geq-1/2\\
{\cal F}\left(1+\alpha\right)\left(-k\eta\right)^{6+2\alpha}k^{-1},\ \alpha\leq-1/2
\end{array}\end{cases}
\]
\begin{equation}
\approx\frac{H^{4}}{2\pi^{2}}\begin{cases}
\begin{array}{c}
{\cal F}\left(\alpha\right)\begin{cases}
\begin{array}{c}
\frac{1}{4-2\alpha}\begin{cases}
\begin{array}{c}
1,\ \alpha<2\\
-\left(\frac{a_{i}}{a_{f}}\right)^{4-2\alpha},\ \alpha>2
\end{array}\end{cases}\\
\log\left(\frac{a_{f}}{a_{i}}\right),\ \alpha=2
\end{array} & ,\alpha\geq-1/2\end{cases}\\
{\cal F}\left(1+\alpha\right)\begin{cases}
\begin{array}{c}
\frac{1}{6+2\alpha}\begin{cases}
\begin{array}{c}
1,\ \alpha>-3\\
-\left(\frac{a_{i}}{a_{f}}\right)^{6+2\alpha},\ \alpha<-3
\end{array}\end{cases}\\
\log\left(\frac{a_{f}}{a_{i}}\right),\ \alpha=-3
\end{array} & ,\alpha\leq-1/2\end{cases}
\end{array}\end{cases}\label{eq:magnetic energy density}
\end{equation}
 where we have used Eqs. (\ref{eq:Magnetic Spectrum Explicit}) and
(\ref{eq:Constant definition for the magnetic spectrum}). Again,
we can see that there are regions where the magnetic field can back react,
more or less in the same way. 

The expressions in Eq. (\ref{eq:magnetic energy density}) also show that the $\alpha=2$ case does not suffer from any backreaction problem. However, there is another problem associated with this case. The coupling function scales as $f\propto a^{\alpha}$ but at the end of inflation, $f\rightarrow1$ in order to recover the classical electromagnetism. Hence, in the beginning of inflation $f=\left(a_{f}/a_{i}\right)^{\alpha}=e^{- \alpha N_{total}}$ for an inflationary duration with $N_{total}$ e-folds which implies that if $\alpha=2$ and $N_{total}=60$, we get $f=e^{-120}$. Assuming that the EM field couples to charged matter in the usual gauge invariant way then by rewriting the Lagrangian in terms of the canonically normalized EM field, one finds that the physical electric charge scales as $e_{\rm phys}=f^{-1}=e^{120}$, leaving the theory in an uncontrollable  strongly coupled regime. This is called the strong coupling problem and introduces another constraint on the coupling function, namely, $f\gtrsim1$ during inflation \cite{Demozzi:2009fu}. 

As a final remark on the magnetic spectrum, we note that in order to explain the current observations of cosmic magnetic fields on Mpc scales while also satisfying the upper bound from CMB constraints, the amplitude of the magnetic spectrum, $\delta_{B}=\left(d\rho_{B}/d\log k\right)^{1/2}$, at the end of inflation, should be between $10^{-9}-10^{-16}$, in Planck units. Any proposed solution to the strong coupling problem should also satisfy this condition.

\section{Sawtooth coupling}

\begin{figure}
\begin{centering}
\includegraphics[scale=0.6]{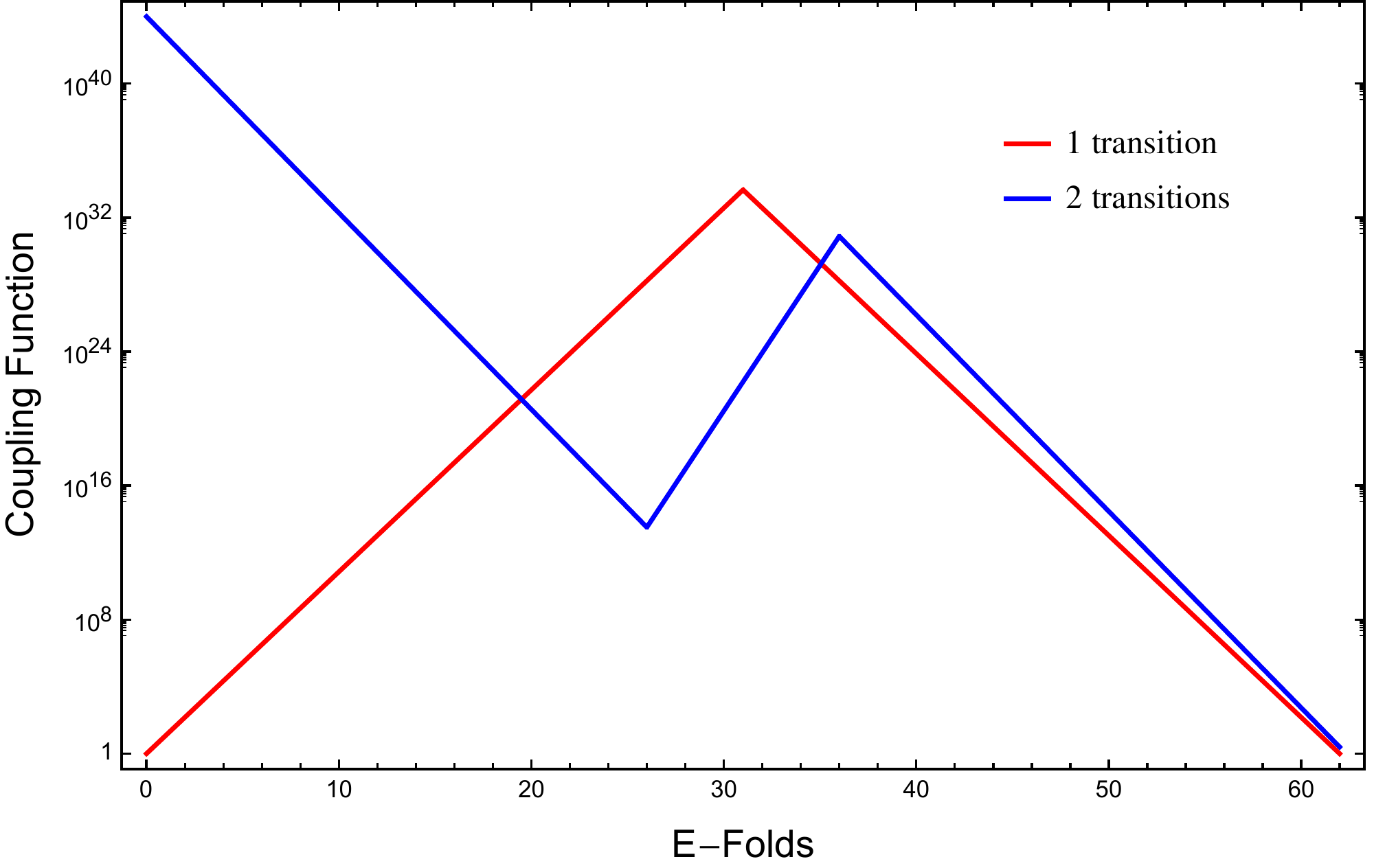}
\par\end{centering}
\caption{A sketch of the coupling function for the cases of 1 and 2 transitions\label{fig:Coupling function}}
\end{figure}

\label{sec:ourmodel}

In this section we analyze the effect on the magnetic spectrum of
having transitions in the coupling function. Related behaviors have
been studied in a different context in Refs. \cite{Starobinsky:1992ts,Leach:2000yw,Leach:2001zf,Jain:2007au}
wherein a sudden break of slow roll conditions leads to an enhancement
of the curvature perturbations on super horizon scales. 

The natural
approach when we allow for transitions would be to simply glue together
the two scale invariant regimes: $\alpha=2$ and $\alpha=-3$. However,
this suffers from two problems. One is the loss in the magnetic spectrum
due to the transition itself, which will be discussed in the next subsection. The other problem is related, again, with either
backreaction or strongly coupled regimes. As we pointed out in the
last section, we can have, at most $13.8$ e-folds of the $\alpha=-3$
behavior. If we glue a stage with $\alpha=2$ we can add $20.7$ e-folds
without entering the strong coupling regimes ($f\ll1$). This means
that we still cannot have more than $34.5$ e-folds of inflation,
which is clearly less than the required number of e-folds of inflation.

Here we will consider coupling functions with one or two transitions
as depicted in Figure \ref{fig:Coupling function}. In each
stage the coupling function has a power law behavior approximately given by $f\propto a^{\alpha_{i}}$.
The coupling function should satisfy $f\gtrsim1$ during inflation
and approach unity in the end thereby avoiding the strongly coupled regimes.

In the following, and in order to keep the notation simple, we assume
that the conformal time is positive definite and goes to zero during
inflation. Also, we will use the term stages to indicate each different
region of the coupling function.

\subsection{One transition}

\label{sec:loss}In the case of one transition we parametrize $f(\eta)$
as

\begin{equation}
f=\begin{cases}
\begin{array}{c}
f_{0}(\frac{\eta_{1}}{\eta})^{\alpha_{1}},\ \eta_{1}<\eta\\
f_{0}(\frac{\eta_{1}}{\eta})^{\alpha_{2}},\ \eta_{1}>\eta
\end{array}\end{cases}
\end{equation}
where $\eta_{1}$ is the time at which the transition occurs and $f_{0}$ is defined by requiring $f(\eta_f)=1$. The
Fourier coefficients of the original vector field are, in
the first stage and after horizon exit, given by

\begin{equation}
A_{1}(k,\eta)=\frac{{\cal A}_{1}(k,\eta)}{f(k,\eta)}=\tilde{C}_{1}(k,\alpha_{1})\left(1+\hat{C}(\alpha_{1})\left(k\eta\right)^{2}\right)+\tilde{D}_{1}(k,\alpha_{1})\eta^{1+2\alpha_{1}},\label{eq:Superhorizon solution for the vector field itself}
\end{equation}
where 
\begin{equation}
\tilde{C}_{1}=\sqrt{\frac{\pi}{4}}\frac{\eta_{1}^{-\alpha_{1}}2^{\alpha_{1}+1/2}k^{-\alpha_{1}-1/2}\exp(i\pi\alpha_{1}/2)}{f_{0}\Gamma(1/2-\alpha_{1})\cos(\pi\alpha_{1})}, \label{eq:C1 tilde}
\end{equation}
\begin{equation}
\tilde{D}_{1}=\sqrt{\frac{\pi}{4}}\frac{\eta_{1}^{-\alpha_{1}}k^{\alpha_{1}+1/2}\exp(i\pi(1-\alpha_{1})/2)}{2^{1/2+\alpha_{1}}f_{0}\Gamma(3/2+\alpha_{1})\cos(\pi\alpha_{1})}
\end{equation}
are coefficients related to the initial conditions and $\hat{C}(\alpha_{i})=-1/(2(1-2\alpha_{i}))$
is a coefficient related to the second order term of the Bessel functions
{[}c.f. Eq. (\ref{eq:Bessel Function Taylor series}){]}. In order
to understand the full solution of $A$, we need to match the super
horizon solutions at each transition. This is similar to matching
the curvature perturbation before and after a bounce on a surface
of constant energy \cite{Deruelle:1995kd,Durrer:2002jn}.

By continuity of the energy stored in both fields, the quantities which have to be matched at the transition points are the vector field $A$ itself and its derivative. In fact, it has been shown that only by matching the vector field itself, and not the canonically normalized field, the underlying EM duality symmetry under the combined exchange of $E \leftrightarrow B$ and inversion of the coupling function $\alpha \rightarrow -\alpha$ is preserved \cite{Brustein:1998kq,Buonanno:1997zk}.

The superhorizon solution for $A$ after each transition is generically
given by 
\begin{equation}
A_{i}(k,\eta)=\tilde C_{i}(k)\left(1+\hat{C}_{i}(\alpha_{i})(k\eta)^{2}\right)+\tilde D_{i}(\alpha_{i})\eta^{1+2\alpha_{i}},
\end{equation}
where $C_{i}$ and $D_{i}$ are the coefficients to determine by matching
$A(\eta)$ and $A'(\eta)$.
For one transition the only interesting scenario which avoids
strongly coupled regimes is when $\alpha_{1}>0>\alpha_{2}$. For simplicity
we consider the case $\alpha_{1}>1/2$ and $\alpha_{2}<-1/2$.
In this situation the vector field after the transition is given by
\begin{equation}
A(k,\eta)=\tilde C_{1}\left(1+\frac{2\hat{C}_{1}(k\eta_{1})^{2}}{1+2\alpha_{2}}\left(\frac{\eta}{\eta_{1}}\right)^{1+2\alpha_{2}}\right).\label{eq:1 Transition - Final expression for A}
\end{equation}

The main feature of having such transitions is the fact that, due
to the matching, the dominant behavior does not pick up immediately
leading to a period where the so-called subdominant solution is
the dominant one\footnote{Another example where the growing solution matches completely to the
decaying solution can be found in the context of string cosmology,
when matching the scalar perturbations from a super-inflationary regime
to a Friedmann era \cite{Brustein:1994kw}.%
}. The duration in e-folds it takes the growing solution to dominate is
\begin{equation}
N_{l}=\frac{-2}{1+2\alpha_{2}}N_{*}-\frac{1}{1+2\alpha_{2}}\ln\left((1+2\alpha_{2})(1-2\alpha_{1})\right),\label{eq:Time of the loss}
\end{equation}
where $N_{*}$ is the number of e-folds that a given mode was outside
the horizon. During this period $N_{l}>0$,
there is a loss/suppression in the spectrum given by, 
\begin{equation}
\left|A\right|\sim {\rm const.}\, \Rightarrow\, \frac{d\rho_{B}}{d\log k}\propto\eta^{4-2\alpha_{2}}.\label{eq:Subdominant Behavior}
\end{equation}
Nevertheless, we can proceed by computing explicitly the final magnetic
spectrum and verify whether we can get an improvement over the case
without transitions. Assuming that the increasing solution picks up
before the end of inflation, $1<(k\eta_{1})^{2}\left(\eta_{f}/\eta_{1}\right)^{1+2\alpha_{2}},$
and using Eqs. (\ref{eq:Magnetic Spectrum}) and (\ref{eq:1 Transition - Final expression for A})
we obtain 
\begin{equation}
\frac{d\rho_{B}(\eta_{f},k)}{d\log k}=\frac{{\cal F}(-\alpha_{1})H^{4}}{2\pi^{2}(1-2\alpha_{1})^{2}(1+2\alpha_{2})^{2}}(k\eta_{f})^{8-2\alpha_{1}}\left(\frac{\eta_{1}}{\eta_{f}}\right)^{2-2\alpha_{1}-2\alpha_{2}}.
\end{equation}
Therefore, an improvement over the case of a monotonic coupling function, given by Eq. (\ref{eq:Magnetic Spectrum Explicit}),
can be translated into the following inequality
\begin{equation}
-(6+2\alpha)N_{h}+\ln \left({\cal F}(1+\alpha)\right)<-(8-2\alpha_{1})N_{h}+2(1-\alpha_{1}-\alpha_{2})N_{2}+\ln \left( \frac{{\cal F}(-\alpha_1)}{(1-2\alpha_{1})^{2}(1+2\alpha_{2})^{2}} \right),
\label{eq:1 transition improvement}
\end{equation}
where $\alpha$ is the minimal negative slope allowed in the case without transitions, $N_{h}=N_{t}-N_{*}$
is the number of e-folds since horizon crossing until the end of inflation,
$N_{t}$ is the total number of e-folds during inflation and $N_{2}$
is the number of e-folds of the last stage.

\begin{figure}
\begin{centering}
\includegraphics[scale=0.6]{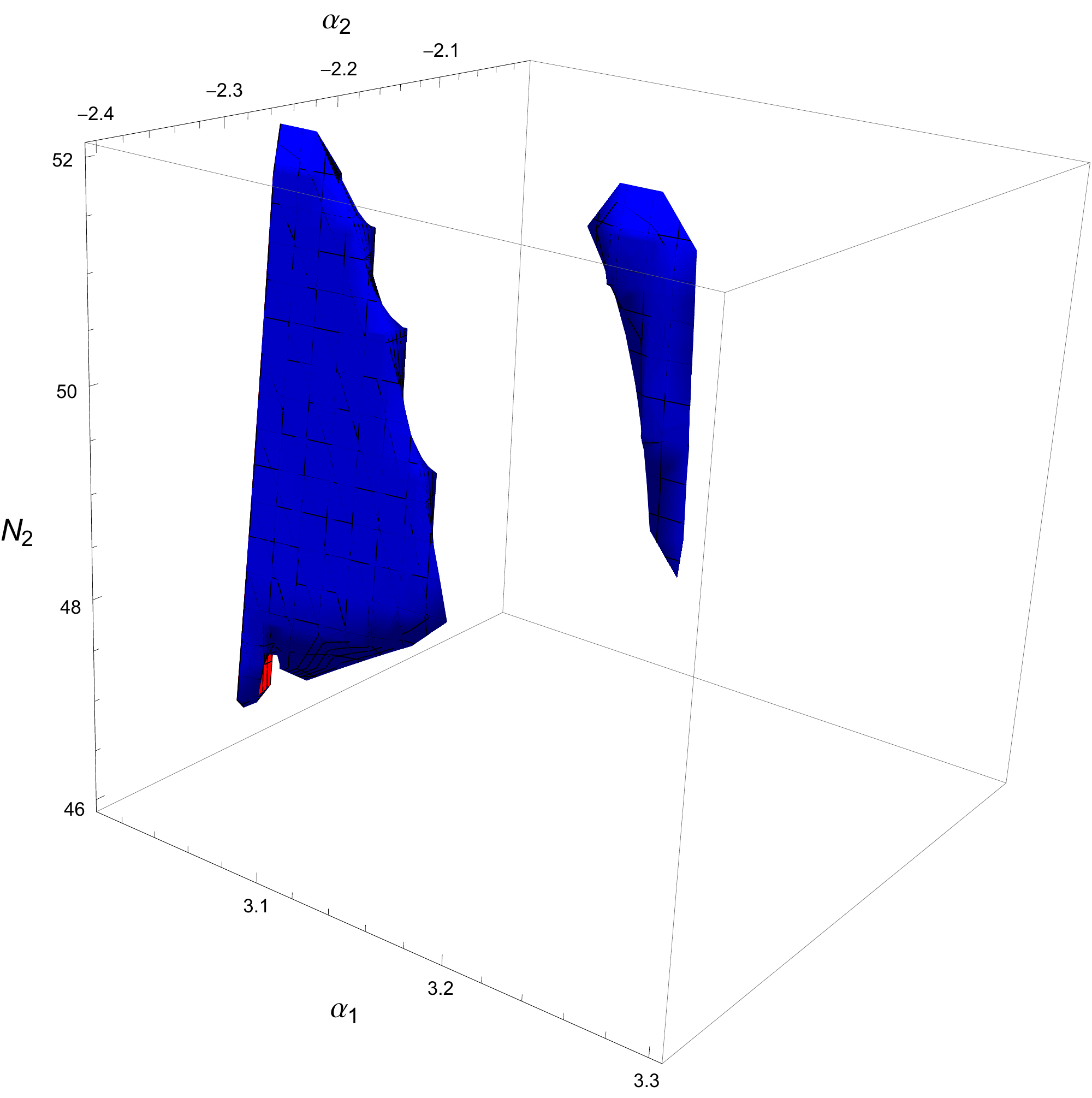}
\par\end{centering}
\caption{The region of parameter space where the magnetic field is more enhanced due to the existence of the transition}
\label{fig:Improved region 1 transition}
\end{figure}

In Fig. \ref{fig:Improved region 1 transition} we show a region of parameters which satisfy inequality (\ref{eq:1 transition improvement}) for an inflation with 60 e-folds and a mode exiting the horizon 8 e-folds after the beginning. Backreaction constrains as well as strongly coupled regimes were taken into account. Although there are solutions,  we did not find any improvement greater than one order of magnitude in the final magnetic field.

\subsection{Two transitions}

In the case of two transitions we parametrize $f(\eta)$ as

\begin{equation}
f=\begin{cases}
\begin{array}{c}
f_{0}(\frac{\eta_{1}}{\eta})^{\alpha_{1}},\ \eta_{1}<\eta\\
f_{0}(\frac{\eta_{1}}{\eta})^{\alpha_{2}},\ \eta_{2}<\eta<\eta_{1}\\
f_{0}(\frac{\eta_{1}}{\eta_{2}})^{\alpha_{2}}(\frac{\eta_{2}}{\eta})^{\alpha_{3}},\ \eta<\eta_{2}
\end{array} & \end{cases}
\end{equation}
where $\eta_{1}$ and $\eta_{2}$ are the times of the transitions. Again, $f_0$ is defined by imposing $f(\eta_f)=1$.
The solution for $A_{1}(k,\eta)$ in the first stage is the same of
Eq. (\ref{eq:Superhorizon solution for the vector field itself}).
We consider, for simplicity, the case where $\alpha_{1}<-1/2$,
$\alpha_{2}>1/2$ and $\alpha_{3}<-1/2$. After performing the
matching at each of the two transitions one obtains the solution for $A$ in
the third stage given by

\baq
A_{3}(k,\eta)&=&\tilde{D}_{1}\eta_{1}^{1+2\alpha_{1}}\left(1-\frac{1+2\alpha_{1}}{1+2\alpha_{2}}\right)\label{eq:Two transitions - Final expression for A} \\
&+&\frac{\tilde{D}_{1}\eta_{1}^{1+2\alpha_{1}}}{1+2\alpha_{3}}\left(\frac{\eta}{\eta_{2}}\right)^{1+2\alpha_{3}}\left(2\left(1-\frac{1+2\alpha_{1}}{1+2\alpha_{2}}\right)(k\eta_{2})^{2}\left(\hat{C}(\alpha_{2})-\hat{C}(\alpha_{3})\right)+(1+2\alpha_{1})\left(\frac{\eta_{2}}{\eta_{1}}\right)^{1+2\alpha_{2}}\right). \nonumber
\eaq
Regarding the final magnetic spectrum, it has two different limits depending on which term on the second line of Eq. (\ref{eq:Two transitions - Final expression for A})
dominates. One regime occurs when $(k\eta_{2})^{2}\gg(\eta_{2}/\eta_{1})^{1+2\alpha_{2}}$. In that case, assuming $1\ll(\eta_{f}/\eta_{2})^{1+2\alpha_{3}}(k\eta_{2})^{2}$,
one gets the following magnetic spectrum at the end of inflation,

\begin{equation}
\frac{d\rho_{B}(\eta_{f},k)}{d\log k}=\frac{16(\alpha_{2}-\alpha_{1})^{2}(\alpha_{2}-\alpha_{3})^{2}{\cal F}(1+\alpha_1)}{2\pi^{2}(1+2\alpha_{3})^{2}(1+2\alpha_{2})^{2}}H^{4}(k\eta_{f})^{10+2\alpha_{1}}\left(\frac{\eta_{f}}{\eta_{2}}\right)^{2(\alpha_{3}-\alpha_{1}-2)}\left(\frac{\eta_{2}}{\eta_{1}}\right)^{-2(\alpha_{1}+\alpha_{2}+1)}.
\end{equation}
An improved result in this situation corresponds to
\begin{equation}
-(6+2\alpha)N_{h}<-(10+2\alpha_{1})N_{h}-2(\alpha_{3}-\alpha_{1}-2)N_{2}+2(\alpha_{1}+\alpha_{2}+1)N_{1},
\end{equation}
where $N_{1}$ is the time duration in e-folds of the middle stage.

When $(k\eta_{2})^{2}\ll(\eta_{2}/\eta_{1})^{1+2\alpha_{2}}$, we are in the other limit and the magnetic spectrum at the end of inflation, assuming $1<(\eta_{2}/\eta_{1})^{1+2\alpha_{2}}\left(\eta_{f}/\eta_{2}\right)^{1+2\alpha_{3}}$,
is given by 
\begin{equation}
\frac{d\rho_{B}(\eta_{f},k)}{d\log k}=\frac{(1+2\alpha_{1})^{2}{\cal F}(1+\alpha_{1})}{2\pi^{2}(1+2\alpha_{3})^{2}}H^{4}(k\eta_{f})^{6+2\alpha_{1}}\left(\frac{\eta_{f}}{\eta_{2}}\right)^{2(\alpha_{3}-\alpha_{1})}\left(\frac{\eta_{2}}{\eta_{1}}\right)^{2(\alpha_{2}-\alpha_{1})}.
\end{equation}
Consequently, the final magnetic spectrum is improved in this latter case whenever 
\begin{equation}
-(6+2\alpha)N_{h}<-(6+2\alpha_{1})N_{h}-2(\alpha_{3}-\alpha_{1})N_{2}-2(\alpha_{2}-\alpha_{1})N_{1}.
\end{equation}
We did not find any region in the parameter space which would lead to an improvement in the final magnetic field strength over the case without transitions.

\subsection{Backreaction on perturbations}

Finally, since the time-derivatives of the coupling $f$ become large at the transition points, one might also worry about possible effects of the slow-roll parameters from back reaction\footnote{We thank Nemanja Kaloper for pointing out this additional constraint to us.}, which would lead to features in the power spectrum or large non-Gaussianity. A detailed analysis would involve calculating the one-loop effective potential for the inflaton, but here we are just interested in an order of magnitude estimate. As an estimate of the correction to the inflaton potential, we write $\Delta V(\phi) \sim f^2(\phi) \left< \rho_{EM}/f^2\right>$. The backreaction constraint that we already considered corresponds to requiring $\Delta V(\phi) \ll V(\phi)$, but as mentioned $\Delta V(\phi)$ will also contribute to the slow-roll parameters. We obtain a correction to the first slow-roll parameter $\epsilon$ given by 
\beq
\Delta \epsilon = \frac{1}{2}\left(\frac{\Delta V_\phi}{V}\right)^2 \sim 2 \left(\frac{f_\phi}{f}\right)^2\left(\frac{\rho_{EM}}{\rho_{total}}\right)^2    
\eeq
and similarly for the second slow-roll parameter $\eta$, 
\beq
\Delta \eta = \frac{\Delta V_{\phi\phi}}{V} \sim 2\left(\left(\frac{f_\phi}{f}\right)^2 +\frac{f_{\phi\phi}}{f}\right) \frac{\rho_{EM}}{\rho_{total}}
\eeq
where sub-index $\phi$ indicates the derivative with respect to $\phi$ and $\rho_{total}$ is the total energy density. Now in order to estimate the derivatives of the coupling function, we use $f_\phi = \frac{1}{\sqrt{2\ep}}f_N$, where sub-index $N$ denotes derivatives with respect to the number of e-folds. A simple estimate gives $f_N \sim \alpha f$ and $f_{NN} \sim \alpha f/\Delta N$ where $\Delta N$ is the duration of the transition. We find that if the transition is smoother than about one e-fold, $\Delta N \gtrsim 1$, the effects on the slow-roll parameters are small. For very sharp transitions, $\Delta N \ll 1$, the effect on $\ep$ is still small, while there is a large effect on $\eta$. In any case, if the first feature in the coupling function is on scales too small to be seen in the CMB spectrum, then if the transition is very sharp, it would only potentially be visible in the matter power spectrum. 

\section{Deflationary magnetogenesis}
\label{sec:reh}

During inflation the order of magnitude of the magnetic field strength at horizon exit is set by the Hubble rate squared, which is very large compared to the magnetic field strength observed today. But due to flux conservation, the magnetic field always decays as $a^{-2}(t)$ after inflation. Thus, in the standard scenario with almost instantaneous reheating, the magnetic fields are washed out by the subsequent expansion of the universe. As we have seen it is hard to compensate this late time redshift dilution by modifying the inflationary part to produce even larger magnetic fields, so one might instead consider modifications of the post-inflationary evolution. Therefore, we consider the possibility of lowering the scale of inflation jointly with the case of a non-minimal reheating scenario, where the inflaton does not decay immediately into radiation, but the universe is instead dominated by a stiff fluid for a short period just after the end of inflation (called deflation \cite{Spokoiny:1993kt}) as it happens in, for example, disformal \cite{Kaloper:2003yf} or quintessence inflation \cite{Peebles:1998qn,Giovannini:2003jw}.  

In the scenarios with low scale inflation there is less redshift after the end of inflation, which helps to minimize the dilution of the magnetic fields in the post inflation era. In single field inflation, the scale of inflation is related to the observed amplitude of the curvature perturbations and the first slow-roll parameter. This constraint can however be avoided with the curvaton mechanism, where the scale of inflation can be decoupled from the amplitude of the curvature perturbation \cite{Enqvist:2001zp,Lyth:2001nq,Moroi:2001ct}. We will therefore treat the scale of inflation as an independent free parameter.

Furthermore, if the universe is dominated by radiation immediately after the end of inflation, the energy density will redshift like the energy density of radiation, and we will have $\rho_I/\rho_r=(a_0/a_f)^4$, where $a_0$ and $a_f$ are the value of the scale factor today and at the end of inflation, respectively, $\rho_I$ is the energy density at the end of inflation, and $\rho_r$ is the energy density of radiation today. If the universe is instead dominated, after the end of inflation, by a fluid with equation of state $w$ until the end of reheating, where the scale factor is $a_{reh}$, and after dominated by radiation, we would have instead $\rho_I/\rho_r=(a_{reh}/a_f)^{3(1+w)}(a_0/a_{reh})^4$. This last identity can also be written as 
\beq
\frac{a_0}{a_f} = \frac{1}{R}\left(\frac{\rho_I}{\rho_r}\right)^{\frac{1}{4}},
\label{eq:R parameter definition}
\eeq
where we define the reheating parameter $R$\footnote{Our definition of reheating parameter $R$ coincides with the parameter $R_{rad}$ defined in \cite{Demozzi:2012wh}.} 
\beq
\log R =\frac{-1+3 w}{4}\log\left(\frac{a_{reh}}{a_f}\right)~,
\eeq
similarly to \cite{Demozzi:2012wh}.

Thus, when one allows for a reheating stage dominated by a stiff fluid one is effectively minimizing the time a given mode spends on super horizon scales after the end of inflation. On the other hand, we will see that $R$ also has a non-trivial effect on the time a mode spends on super horizon scales during inflation, which goes in the other direction. Below we will show that the combined effect leads to higher values of the present magnetic field.

We start by deriving a generic expression for the present magnetic field\footnote{A less systematic study of related effects has been done in \cite{Martin:2007ue,Demozzi:2012wh}, and our results agree when comparison is possible.}, for a given mode, as a function of the Hubble constant during inflation ($H_{I}$), the reheating parameter ($R$) and the exponent associated with the coupling function ($\alpha$). Then, by maximizing and minimizing, respectively, $R$ and $\alpha$ as a function of $H_{I}$ we derive an upper value on the magnetic field today as a function of $H_{I}$.

\subsection{Magnetic field today}

In order to solve the horizon problem the largest observable scale today $\lambda(t_{0})=H_{0}^{-1}$ should be inside the horizon during inflation, $\lambda(t_{i})<H_{I}^{-1}$. This implies that
\begin{equation}
H_{0}^{-1}\frac{a_{f}}{a_{0}}\frac{a_{i}}{a_{f}}<H_{I}^{-1},\label{eq:Horizon Problem}
\end{equation}
where $a_{i}$ is the scale factor at the beginning of inflation. This inequality can be translated into 
\begin{equation}
N_{t}>\ln\left(\frac{a_{f}}{a_{0}}\right)+\ln\left(\frac{H_{I}}{H_{0}}\right),\label{eq:Minimum e-folds required}
\end{equation}
where $N_{t}$ is the total number of e-folds during inflation. Now, we rewrite $\rho_{r}=\rho_{0}\Omega_{r}$, where $\rho_{0}=3H_{0}^{2}M_{p}^{2}$ is the critical energy density
today and $\Omega_{r}$ is the present radiation density parameter.  By assuming the minimum amount of inflation required and using Eq. (\ref{eq:R parameter definition}) we can write
\begin{equation}
N_{t}=\ln(R)+\frac{1}{2}\ln\left(\frac{H_{I}}{H_{0}}\right)+\frac{1}{4}\ln(\Omega_{r}),
\label{eq:Minimal amount of inflation}
\end{equation}
where we used $\rho_{I}=3H_{I}^{2}M_{p}^{2}$. For instantaneous reheating ($R=1$), $H_I=10^{-6} M_p$, and using the present
cosmological parameters, $\rho_{0}\sim10^{-120}M_{p}^{4}$,
$\Omega_{r}\sim2\times10^{-5}h^{-2}$ and $h=0.7$, it corresponds to $N_{t}\sim60$ e-folds.

In order to compute the final magnetic spectrum we will assume the standard scenario of a monotonic decreasing
coupling function during inflation. As the conformality is restored in the EM action after the end of inflation, the magnetic spectrum evolves after that point as $d\rho_{B}/d\log k\propto a^{-4}$. Using Eqs.(\ref{eq:Magnetic Spectrum}) and (\ref{eq:R parameter definition}), we obtain the following expression for the magnetic spectrum at present time,
\begin{equation}
\left.\frac{d\rho_{B}}{d\log k}\right|_{a_{0}}=\left.\frac{d\rho_{B}}{d\log k}\right|_{a_{f}}\left(\frac{a_{f}}{a_{0}}\right)^{4}=\frac{{\cal F}(1+\alpha)}{2\pi^{2}}H_{I}^{4}\,e^{-N_{h}(6+2\alpha)}R^{4}\frac{\Omega_{r}\rho_{0}}{\rho_{I}}.
\end{equation}
After rewriting $N_{h}=N_{t}-N_{*}$ we can use Eq. (\ref{eq:Minimal amount of inflation}) to get 
\begin{equation}
\left.\frac{d\rho_{B}}{d\log k}\right|_{a_{0}}=\frac{{\cal F}(1+\alpha)}{2\pi^{2}}H_{I}^{4}\,e^{N_{*}(6+2\alpha)}\left(\frac{H_{I}}{H_{0}}\right)^{-(5+\alpha)}\left(R\,\Omega_{r}^{1/4}\right)^{-2(1+\alpha)}.
\label{eq: generic equation for magnetic field today}
\end{equation}
This equation gives the present magnetic field strength for a given mode as a function of $H_{I}$, $R$ and $\alpha$ where we have just assumed
that inflation lasted the minimum amount of time possible. However, we are looking for maximal values of the magnetic field. In order
to maximize $R$ we use an equivalent definition for $R$,
\begin{equation}
\log R=\frac{1-3w}{12(1+w)}\log\left(\frac{\rho_{reh}}{\rho_{I}}\right)<-\frac{1}{12}\log\left(\frac{\rho_{reh}}{\rho_{I}}\right),
\end{equation}
where $\rho_{reh}$ is the energy density at the end of reheating
and $-1/3<w<1$ is the mean equation of state parameter. But $\rho_{reh}$
is also bounded by inflation from above and by nucleosynthesis from below,
\begin{equation}
3\times 10^{-82}M_p^4<\rho_{nucl}<\rho_{reh}<\rho_{I}<10^{-10}M_{p}^{4}.
\end{equation}
This bounds can be translated into the reheating parameter as
\begin{equation}
\log R<-\frac{1}{12}\log\left(\frac{3\times 10^{-82}M_p^4}{3H_I^{2}M_{p}^{2}}\right)\approx-\frac{1}{6}\log\left(10^{-41}\frac{M_{p}}{H_I}\right),
\end{equation}
which implies an upper bound on the value of $R$ as
\begin{equation}
R<10^{41/6}\left(\frac{H_I}{M_{p}}\right)^{1/6}.
\label{1stconstrainonR}
\end{equation}

A similar optimization can be done for $\alpha$. Using the expressions obtained for the backreaction\footnote{While the second version of this draft was written up the paper \cite{Fujita:2013qxa} appeared, where the authors also considered the constraint on $\alpha$ from non-Gaussianity. In the curvaton case with $\rho_I^{1/4}< 10^{14}$ GeV, these new constraints are not any stronger than the backreaction constraint considered here, although in single field inflation the constraints derived in  \cite{Fujita:2013qxa} are very strong, even for low scale inflation. Since scenarios of low-scale inflation and prolonged preheating are however naturally encompassed in the curvaton model, we consider it sufficient to consider the backreaction constraint when deriving an upper bound on the magnetic fields created during inflation in models with $\rho_I^{1/4}< 10^{14}$ GeV.} in
the case of interest, $\alpha<-2$, we can write the minimal value
of $\tilde{\alpha}$ allowed without backreaction as,
\begin{equation}
\tilde{\alpha}=\frac{1}{N_{t}}\ln\left(\frac{H_{I}}{M_{p}}\right)-2,\label{eq:minimal value of alpha}
\end{equation}
which, using Eq. (\ref{eq:Minimum e-folds required}), can be rewritten
as 
\begin{equation}
\tilde{\alpha}=-2+\frac{\ln\left(\frac{H_{I}}{M_{p}}\right)}{\frac{1}{4}\ln\left(\Omega_{r}\frac{H_{I}^{2}}{H_{0}^{2}}\tilde R^{4}\right)}.\label{eq:minimal value of alpha for maximal value of R}
\end{equation}
A further
optimization can be done if one considers that the conformal breaking occurs not at the beginning of inflation but later, although before the scale of interest leaves the horizon. In that case, we would have, maximally,
$N_{t}-N_{*}$ instead of
$N_{t}$ in Eq. (\ref{eq:minimal value of alpha}).

\subsection{Backreaction after inflation}

\begin{figure}
\begin{centering}
\includegraphics[scale=0.6]{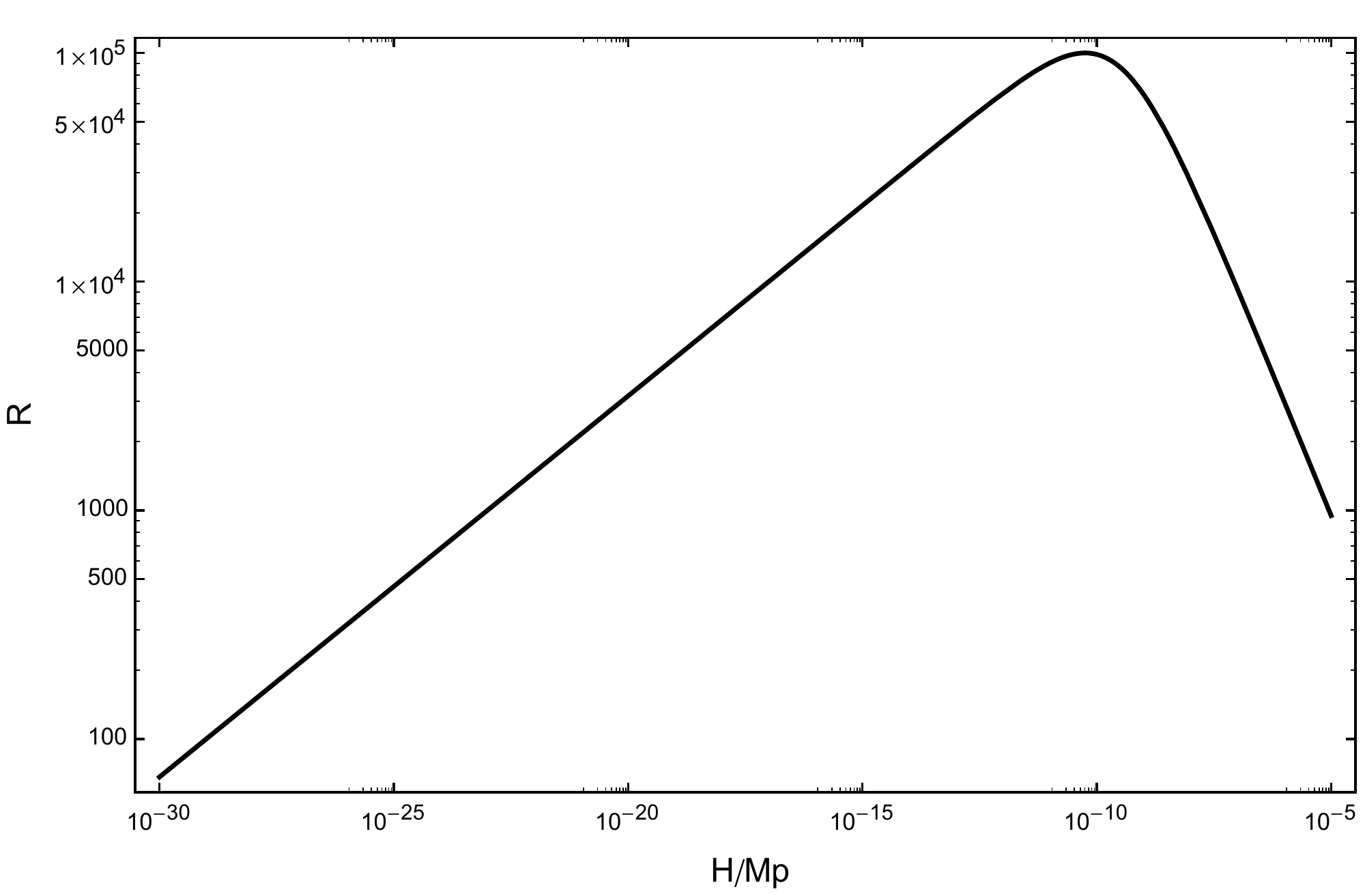}
\par\end{centering}
\caption{Maximal value of $R$ as a function of the Hubble parameter during inflation.}
\label{fig:R max}
\end{figure}

In all the previous sections we ensured that the EM fields are not backreacting during the inflationary dynamics. Nevertheless, we also have to ensure that they do not affect the subsequent dynamics of the universe. Given that in our optimal scenario the fluid which dominates the energy content during the reheating stage has an energy density which decays as $\rho \propto a^{-6}$ and knowing that the energy in the EM fields decay as $\rho_{EM} \propto a^{-4}$ the EM field could, in principle, become dominant rapidly.

\begin{figure}
\begin{centering}
\includegraphics[scale=0.6]{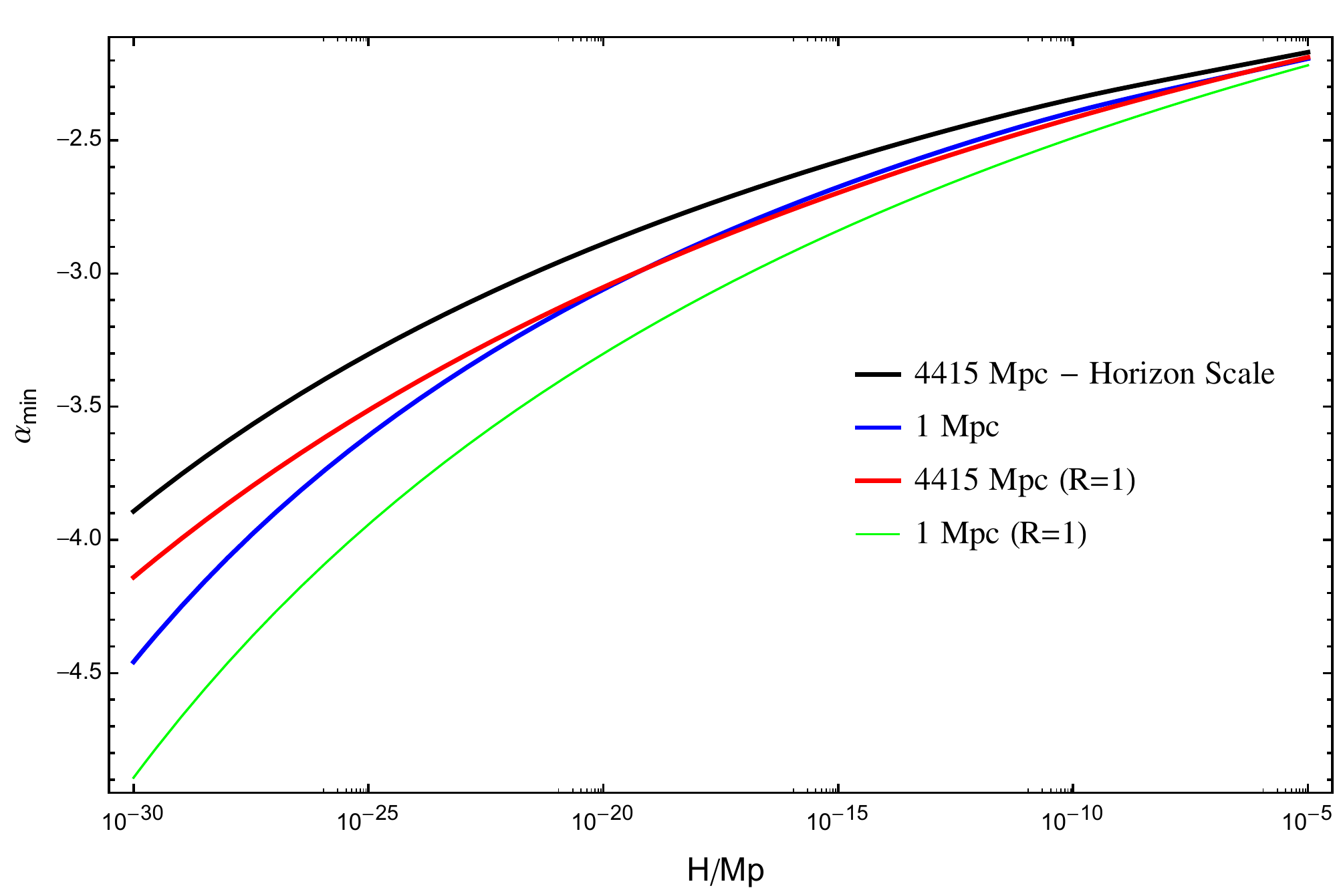}
\par\end{centering}
\caption{Minimal value of $\alpha$ allowed by backreaction and assuming the maximum value for $R$ (blue and black line) or instantaneous reheating (red and green line). The different lines represent different times at which the conformal coupling was broken labeled by the mode which exited the horizon at that moment.}
\label{fig:alpha min}
\end{figure}

In the scenario studied here the energy of the EM field is mainly stored in the electric component because $\alpha<-2$. If one assumes the existence of a preheating stage (see Ref. \cite{Martin:2007ue}) one can show that the magnetic fields do not change in this period but on the other hand, the production of charged particles increase the conductivity of the universe abruptly and screen the electric field by transferring its energy into the plasma of charged particles. In  \cite{Martin:2007ue} it was assumed that the energy density coming from the electric field can be neglected subsequently. But although the electric fields are screened the plasma of the produced charged particles also has to decay as fast as the background energy density in order not to backreact. That requires some extra assumption that we can have a plasma of charged particles behaving as a stiff fluid, as has been suggested to be the case for a fluid of particles with repulsive self-interactions and dominated by the self-interaction energy density \cite{Brisudova:2001ur, Stiele:2010xz}. In Section \ref{results} we present the results assuming the validity of this possibility but we also compare those results with the case where we have instantaneous reheating (without  the stiff phase) and we just lower the scale of inflation.

Regarding the possible backreaction of the magnetic fields after inflation we can do the analysis by looking at our present sky. Since after the end of reheating the energy density of the photons and our magnetic fields evolve in the same way we can write,
\beq
\frac{\rho_B(t_{reh})}{\rho_{reh}}=\frac{\rho_B(t_0)}{\rho_r}<\gamma,
\label{backafterinf}
\eeq
where for $\gamma \ll 1$ in the last inequality implies that the magnetic fields have to be subdominant compared to the total radiation energy density today. Using the cosmological values above and taking $\gamma =1$, we have $\rho_r\sim3.7\times 10^{-12} G^2$ which tells us that the magnetic field today cannot be stronger than $10^{-6}G$. 
As one can see in Figs. (\ref{fig:final magnetic field 1}) and (\ref{fig:final magnetic field 2}) the values obtained for the magnetic field are much lower than this upper limit. However, since these figures only show a restricted range of scales, we should perform a more general analysis.
If we trace Eq. (\ref{backafterinf}) back in time we get, for $\alpha>-3$ and using Eq. (\ref{eq:magnetic energy density}),
\beq
\frac{{\cal F} (1+\alpha)}{24 \pi^2 (3+\alpha)}\left(\frac{H_I}{M_p}\right)^{2}\left(\frac{a_{reh}}{a_f}\right)^2\sim\frac{1}{100}\left(\frac{H_I}{M_p}\right)^{2}R^4<\gamma.
\label{2ndconstrainonR}
\eeq
This is a non-trivial constraint in our analysis which we have to take into account when defining the maximal value of $R$. Instead, for $\alpha<-3$ we get
\beq
\frac{{\cal F} (1+\alpha)}{24 \pi^2 (3+\alpha)}\left(\frac{H_I}{M_p}\right)^{2}\left(\frac{a_{reh}}{a_f}\right)^2e^{-N_h(6+2\alpha)}\sim
\frac{1}{100}\left(\frac{H_I}{M_p}\right)^{2}R^4\left(R\left(\frac{H_I}{H_0}\right)^{1/2}\Omega_r^{1/4}\right)^{-(6+2\alpha)}e^{N_*(6+2\alpha)}<\gamma.
\eeq
which is trivially satisfied for any reasonable value of $\gamma$.

The two non-trivial constraints on $R$, Eqs. (\ref{1stconstrainonR}) and (\ref{2ndconstrainonR}) can be combined into one single constraint, which defines a maximal $R$ as a function of $H_I$, given by
\beq \label{mastereq}
\tilde R \sim \left( \frac{1}{3 \gamma^{1/4}} \left(\frac{H_I}{M_p}\right)^{1/2} + 10^{-41/6}\left(\frac{H_I}{M_p}\right)^{-1/6} \right)^{-1}.
\eeq

\begin{figure}
\begin{centering}
\includegraphics[scale=0.6]{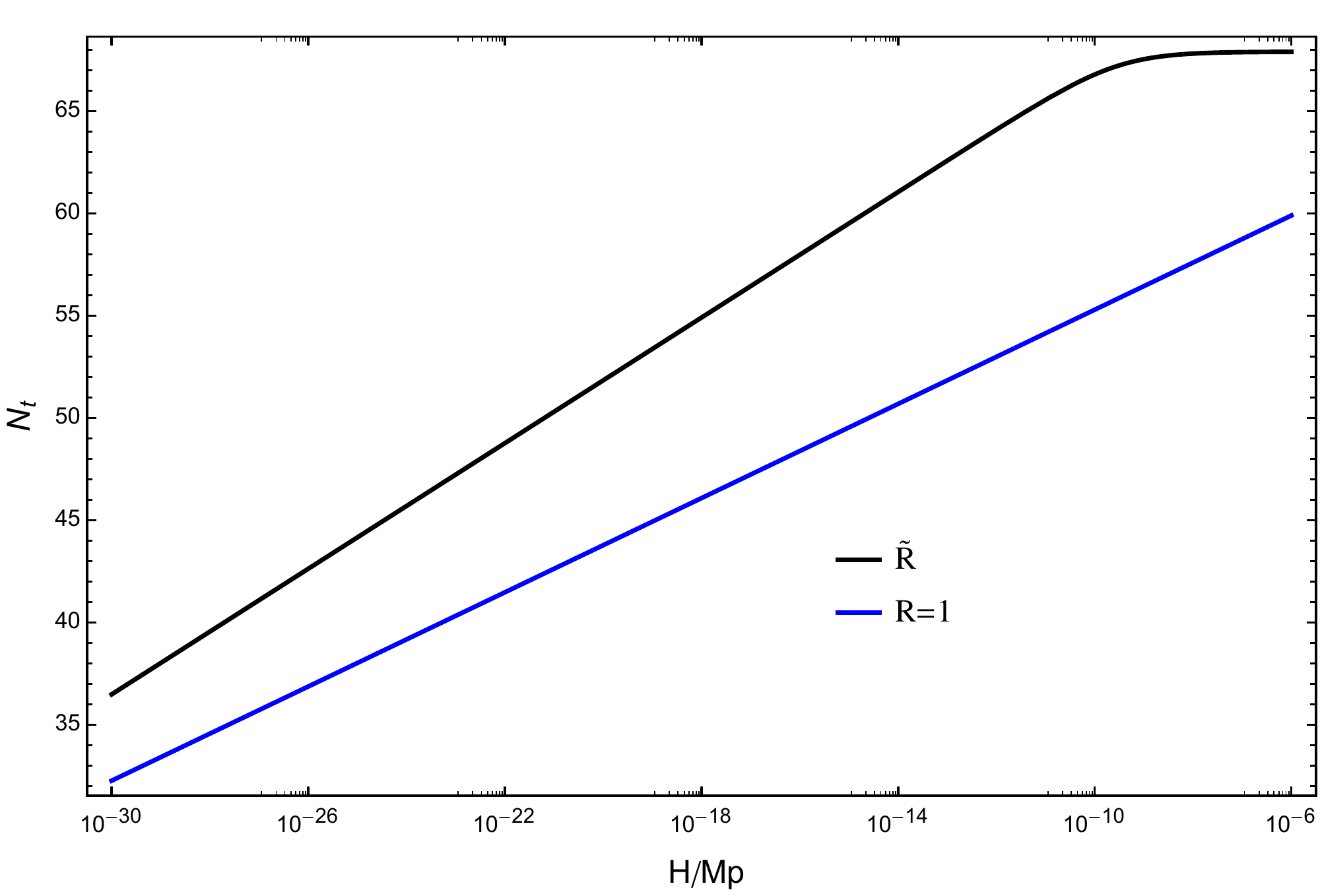}
\par\end{centering}
\caption{Total number of e-folds during inflation in the optimal scenario for deflationary magnetogenesis compared to the scenario of instantaneous reheating, both as a function of $H_I$.}
\label{fig:Ntotal}
\end{figure}

In Fig. (\ref{fig:R max}) we show the relation between $\tilde R$ and $H_I$ for $\gamma =1$.
Using the definition of $\tilde R$ we plot, in Fig. (\ref{fig:alpha min}), the value of $\tilde \alpha$ given by Eq. (\ref{eq:minimal value of alpha for maximal value of R}) and also the optimal value of $\alpha$ allowed for instantaneous reheating, both for different times of conformal breaking and $\gamma =1$. It is interesting to see that the scale invariant scenario, $\alpha=-3$, is allowed for $H_I\sim10^{-22}M_p$ or even higher values of $H_I$ in the case of a delayed conformal breaking. We also plot in Fig. (\ref{fig:Ntotal}) the corresponding total number of e-folds during inflation as a function of $H_I$.

Finally, we note that in the case where the constraint in (\ref{2ndconstrainonR}) is saturated, we would have to worry about the magnetic fields effectively behaving similar to a curvaton \cite{Enqvist:2001zp,Lyth:2001nq,Moroi:2001ct}, and possible large non-Gaussian curvature perturbations being sourced by the non-adiabatic pressure of the magnetic fluid. One can see from Fig. (\ref{fig:alpha min}) that this only becomes relevant for $\tilde\alpha> -2.5$. On the other hand the induced power spectrum of curvature perturbations will scale as $(k/(a_f H_I))^{12+4\alpha}$ in this regime \cite{Caprini:2009vk}, and are therefore strongly suppressed on CMB scales. The constraint from black hole formation on small scales does however still apply \cite{Josan:2009qn}, and hence we need to have $\gamma \ll 10^{-2}$ in order to satisfy this constraint. This will however not change our bound in (\ref{mastereq}) appreciably.

\subsection{Results} \label{results}

After writing all the optimal parameters as a function of $H_{I}$ we can write the final expression for the upper value of the magnetic field today, $B_{0}=(d\rho_{B}/d\log k)^{1/2}$ as,

\begin{equation}
B_{0}=\left(\frac{{\cal F}(1+\tilde \alpha)}{2\pi^{2}}\right)^{1/2}H_{I}^{2}e^{N_{*}(3+\tilde \alpha)}\left(\frac{H_{I}}{H_{0}}\right)^{-\frac{1}{2}(5+\tilde \alpha)}\left(\tilde R\Omega_{r}^{1/4}\right)^{-(1+\tilde \alpha)}.
\label{eq: final maximal B0}
\end{equation}

\begin{figure}
\begin{centering}
\includegraphics[scale=0.6]{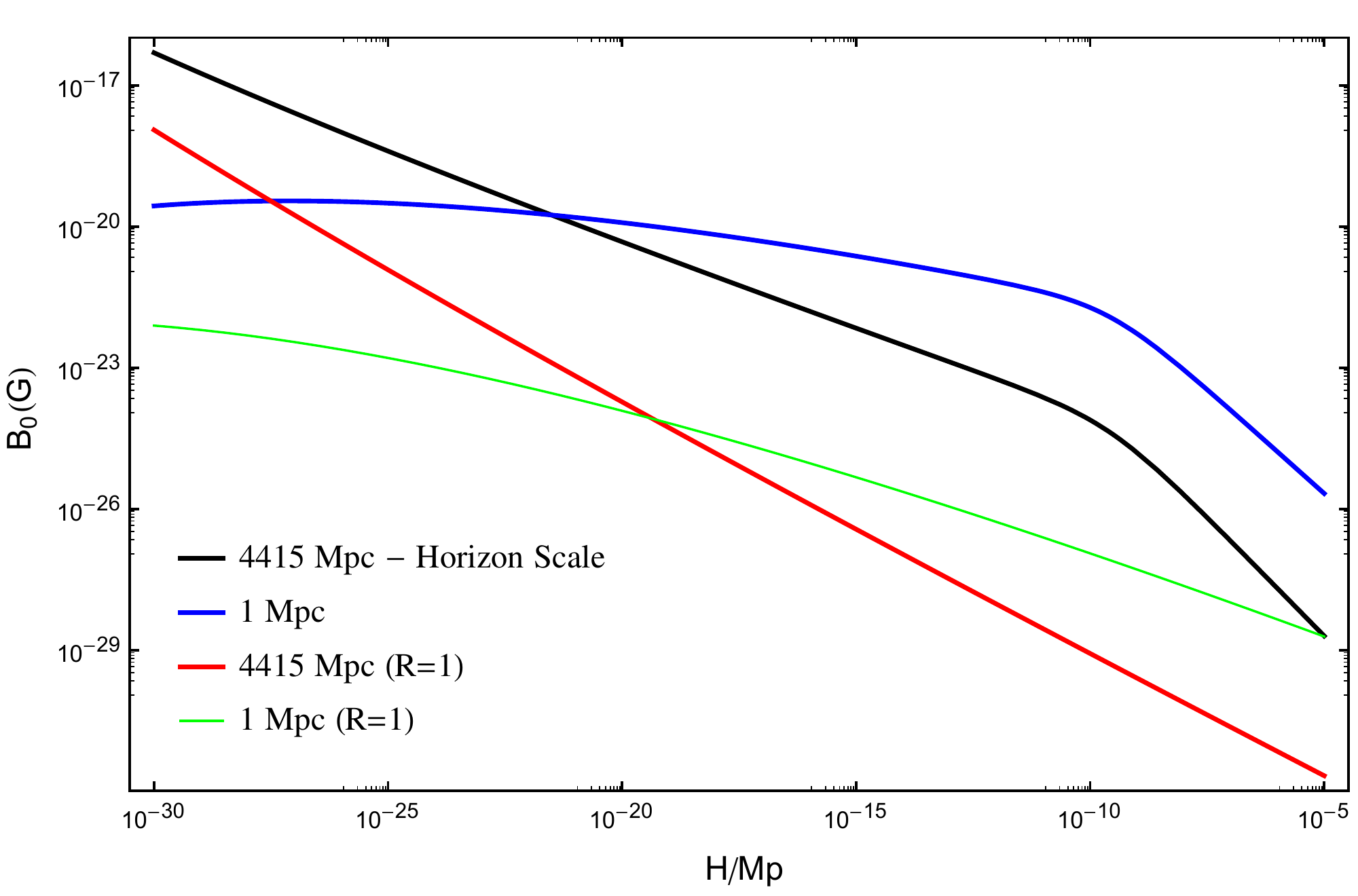}
\par\end{centering}
\caption{Maximal present magnetic field assuming the maximum value for $R$ (blue and black line) or instantaneous reheating (red and green line) and the lower value for $\alpha$, as a function of the Hubble parameter during inflation $H_{I}$. The different lines correspond to different scales.}
\label{fig:final magnetic field 1}
\end{figure}

In Fig. (\ref{fig:final magnetic field 1}) we present the upper value for $B_{0}$ as a function of the Hubble parameter during inflation $H_{I}$ for the horizon and the $Mpc$ scale and we compare those results with the upper value for $B_{0}$ in the case of instantaneous reheating, for the same scales. First, it is very interesting to notice the existence of turning points at $H_{I}\sim10^{-22}M_p$ and $H_{I}\sim10^{-20}M_p$, corresponding to $\alpha=-3$, which separates the red spectrum from blue spectrum. At the $Mpc$ scale, we can see that $10^{-20}G$ can be generated at low energy scales, $H_{I}= 10^{-30} M_p$, decreasing this value down to $10^{-26}G$ at $H_{I}=10^{-5}M_{p}$. Due to the fact that the spectrum is blue shifted for $H<10^{-22}M_{p}$ it is possible to generate $10^{-17}G$ magnetic fields at horizon scales for a TeV inflation scale. When the reheating is instantaneous, these values decrease by $2$ orders of magnitude for $H_{I}= 10^{-30} M_p$ and by 3 orders of magnitude for $H_{I}=10^{-5}M_{p}$. The reason is that in the optimal scenario for low scale inflation, the reheating parameter is small anyway.

\begin{figure}
\begin{centering}
\includegraphics[scale=0.6]{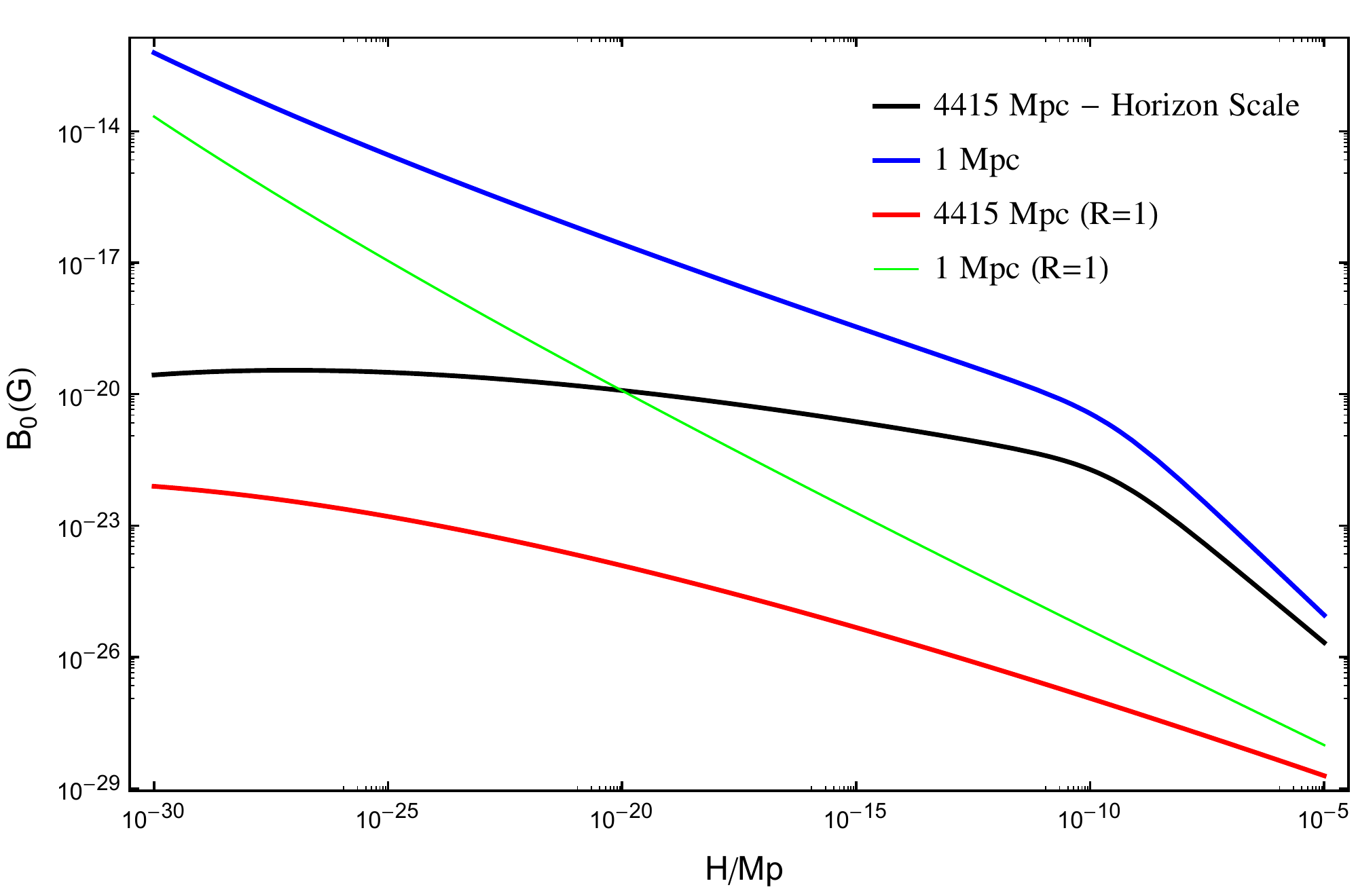}
\par\end{centering}
\caption{Present magnetic field at $Mpc$ scale assuming the maximum value for $R$ (blue and black line) or instantaneous reheating (red and green line), and the lower value for $\alpha$, as a function of the Hubble parameter during inflation $H_{I}$. The different lines correspond to different scales at which the conformal invariance was broken labeled by the mode which exited the horizon at that time.}
\label{fig:final magnetic field 2}
\end{figure}

In Fig. (\ref{fig:final magnetic field 2}) we plot the same quantities only for the $Mpc$ scale but we compare to the case where the breaking of conformal invariance occurred around the time the $Mpc$ scale left the horizon.
We can see that stronger magnetic fields are generated when the breaking of conformal invariance occurs closer to the horizon exit of the $Mpc$ scale. This is easy to understand in the sense that by breaking the conformal invariance later one allows the value of $\alpha$ to be smaller. 
The maximum value of the magnetic fields allowed goes from $10^{-13} G$ at $H_{I}=10^{-30}M_{p}$ to $10^{-25} G$ at $H_{I}=10^{-5}M_{p}$ and even in the case of instantaneous reheating one has, in that situation, $10^{-14} G$ at $H_{I}=10^{-30}M_{p}$ although $10^{-29} G$ at $H_{I}=10^{-5}M_{p}$.

\section{Conclusion}  \label{sec:conclusion}

In this article we have considered two different approaches to optimize the generation and subsequent evolution of primordial magnetic fields from inflation. 

In the first scenario we consider a new model of inflationary magnetogenesis based on transitions in the coupling function during inflation. These transitions connect multiple stages of the coupling function with different power law behaviors. The main consequence of such a type of coupling function is the appearance of losses in the spectrum due to the matching at the transition point, where the growing mode before the transition is matched almost completely to the decaying mode after the transition. Although it is a self-consistent inflationary model for the generation of large scale magnetic fields which solves the strong coupling problem and avoids backreaction, the generated magnetic fields are only enhanced at most one order of magnitude more than in the case without transitions. This result was found in the case where we have one transition. In the case of two transitions we did not find any self-consistent improvement.

In our second approach, deflationary magnetogenesis, much larger magnetic fields can be obtained. We did a study of the consequences of lowering the energy scale of inflation as well as of the existence of a prolonged reheating stage on the primordial magnetic fields. Without specifying the details of the reheating we derived an equation which gives the present magnetic field as a function of the Hubble parameter during inflation, $H_I$, the exponent of the coupling function which breaks the conformal invariance, $\alpha$, the reheating parameter, $R$, and some known cosmological parameters. Using this we derived a scale dependent upper bound for the present magnetic field only as a function of the Hubble parameter during inflation by requiring the optimal values of $\alpha$ and $R$. We found that at $Mpc$ scale the maximal magnetic field decreases with the energy scale of inflation going from $10^{-20} G$ at $H_{I}=10^{-30}M_{p}$ to $10^{-26}G$ at $H_{I}=10^{-5} M_{p}$. Interestingly, at horizon scale the maximal value can go up to $10^{-17}G$ at $H_{I}=10^{-30}M_{P}$. In the case of instantaneous reheating these values decrease between 2 and 3 orders of magnitude. Furthermore, if one assumes that the breaking of conformal invariance occurs only around the time the $Mpc$ scale left the horizon during inflation, then the upper value on the maximal magnetic field goes from $10^{-13} G$ to $10^{-25}G$ for the same range of values of $H_{I}$. Even in the case of instantaneous reheating (without the stiff fluid phase) we found an upper value of $10^{-14} G$ at TeV scale inflation. These are the main results of our work.

We would like to note that since the coupling function is a piecewise monomial function of the inflaton in all of the models studied here, for each section the magnetic consistency relation derived previously in \cite{Caldwell:2011ra,Motta:2012rn,Jain:2012ga,Jain:2012vm,Biagetti:2013qqa}, can be used to compute the cross-correlation function of the magnetic field with the curvature perturbation.

In \cite{Fujita:2012rb} an upper limit on the strength of magnetic fields from inflation was derived. It can be checked that for $R\gg1$, the bounds in \cite{Fujita:2012rb} are weakened, and are in any case consistent with the upper bounds derived here from deflationary magnetogenesis.

While our results  shows that breaking conformal invariance during inflation might still not be excluded as an explanation for the recent claims of $10^{-16}G$ magnetic fields on $Mpc$ scales and the seed fields for the galactic dynamo, it would be interesting to study concrete models where this is realized, in order to further explore this possibility. This is however beyond the scope of the present work, and we have left it for the future to investigate more concrete models.

\acknowledgments

We would like to thank Ruth Durrer, Tomohiro Fujita, Nemanja Kaloper, Marco Peloso and Antonio Riotto for useful comments. We are also grateful to Tomohiro Fujita for pointing out an issue with the numerical solutions in the first version of this paper. RKJ is supported by an individual postdoctoral fellowship from the Danish council for independent research in Natural Sciences. MSS is supported by a Jr. Group Leader Fellowship from the Lundbeck Foundation. The CP$^3$-Origins centre is partially funded by the Danish National Research Foundation, grant number DNRF90.


\bibliographystyle{JHEP}

\bibliography{ReferencesFor_OnTheStrongCouplProblm}

\end{document}